\documentclass[12pt]{iopart}

\expandafter\let\csname equation*\endcsname\relax
\expandafter\let\csname endequation*\endcsname\relax

\usepackage{amsfonts,amssymb,graphicx,hyperref,epstopdf,color,tikz,scalerel,amsmath,braket,enumerate,array,mathptmx}
\usepackage{iopams}
\usepackage[T1]{fontenc}
\usepackage[utf8]{inputenc}
\hypersetup{colorlinks,citecolor=blue,filecolor=blue,linkcolor=blue,urlcolor=blue}
\usetikzlibrary{svg.path}\definecolor{orcidlogocol}{HTML}{A6CE39}
\tikzset{orcidlogo/.pic={\fill[orcidlogocol] svg{M256,128c0,70.7-57.3,128-128,128C57.3,256,0,198.7,0,128C0,57.3,57.3,0,128,0C198.7,0,256,57.3,256,128z};
\fill[white] svg{M86.3,186.2H70.9V79.1h15.4v48.4V186.2z}svg{M108.9,79.1h41.6c39.6,0,57,28.3,57,53.6c0,27.5-21.5,53.6-56.8,53.6h-41.8V79.1z M124.3,172.4h24.5c34.9,0,42.9-26.5,42.9-39.7c0-21.5-13.7-39.7-43.7-39.7h-23.7V172.4z}
svg{M88.7,56.8c0,5.5-4.5,10.1-10.1,10.1c-5.6,0-10.1-4.6-10.1-10.1c0-5.6,4.5-10.1,10.1-10.1C84.2,46.7,88.7,51.3,88.7,56.8z};
}}
\newcommand\orcidicon[1]{\href{https://orcid.org/#1}{\mbox{\scalerel*{
\begin{tikzpicture}[yscale=-1,transform shape] \pic{orcidlogo}; \end{tikzpicture}}{|}}}} 

\usepackage[numbers,sort&compress]{natbib}

\newcommand{\be}{\begin {equation}}
\newcommand{\ee}{\end {equation}}
\newcommand{\beqa}{\begin {eqnarray}}
\newcommand{\eeqa}{\end {eqnarray}}

\begin{document}

\title[Regulating the higher harmonic cutoffs via sinc pulse]{Regulating the higher harmonic cutoffs via sinc pulse}

\author{Rambabu Rajpoot \footnote[1]{ramrajpoot3@gmail.com}, Amol R. Holkundkar \footnote[2]{amol.holkundkar@pilani.bits-pilani.ac.in}, and Jayendra N. Bandyopadhyay \footnote[3]{jnbandyo@gmail.com}}


\vspace{3pt}
\address{Department of Physics, Birla Institute of Technology and Science - Pilani, Rajasthan, 333031, India.}

\vspace{10pt}
\begin{indented}
\item[]\today
\end{indented}

\begin{abstract}
 
We theoretically investigate the generation of higher harmonics and the construction of a single attosecond pulse by means of two oppositely polarized sinc-shaped driver pulses. In comparison to a few-cycle Gaussian pulse of the same energy, here we observe a significant broadening in the bandwidth of an XUV/soft-Xray supercontinuum spectrum in the synthesized pulse. Furthermore, we observe that the harmonic cutoff and its corresponding intensity follow a well-defined scaling with the {\it delay parameter} between the two pulses. In principle, this delay can easily be tuned in on an optical bench. The typical nature of the synthesized pulse ensures the generation of single attosecond pulse instead of a pulse train. In this case, we obtain a single attosecond pulse of duration $\sim 27$ attosecond in XUV/soft-Xray regime of the electromagnetic spectrum. Depending on the delay parameter we observe an enhancement in some satellite harmonics. The proposed setup promises a highly tunable source of energetic photons, wherein the energy of the photons can easily be controlled from XUV to soft-Xray regime by simply changing the delay between two oppositely polarized sinc-pulses.
 
\end{abstract}
\maketitle
 
\section{Introduction}

The last decade has witnessed rapid development in the field of higher harmonic generation (HHG), both in experimental and theoretical front \cite{Ciappina_2017}. The HHG and, consequently the production of attosecond pulses (ASPs) have enabled researchers to probe the fundamental processes of atomic and molecular phenomenon with unprecedented resolution. The ASP is particularly important for investigating the electron correlation effects and observing characteristic temporal delays in photoemission from different atomic orbitals \cite{PhysRevLett.110.023003,PhysRevLett.115.153001}. The ASP can also probe detailed microscopic motion of electrons in atoms, molecules, or any other nanoscale structures, and those effectively bridge the gap between various fields of basic sciences \cite{Chini2014_nat, Krausz2009_RMP, Corkum2007_NatPhy, Paul1689_Science, Calegari2016_JPB}.

The HHG originates from the interaction of intense laser with gas atoms, which leads to the generation of coherent radiation at higher harmonics of the laser frequency. Experimentally, it has been observed that Higher Harmonic spectrum consists of a plateau where harmonic intensity is nearly constant over many orders of magnitude followed by a sharp cutoff \cite{Corkum1993_PRL, Lewenstein1994_PRA}. Ever since the inception of the idea of the generation of the higher harmonics by laser-atom interaction, the research objectives around the globe are aimed towards the enhancement of the harmonic cutoff of the HHG and also to increase the corresponding intensity of the emitted harmonics. There are various studies devoted to achieve these goals. The effect of the pulse chirp \cite{Liu2019_OptComm, Peng2018_PRA, Feng2017_MolPhys, Liu2019_LaserPhys, Han2019OpticExp, Lara2016_PRL}, pulse duration \cite{Zhou1996_PRL, Ganeev2013_JOSAB, Chang2019_OSACont}, synthesis of laser pulse using two or more colors laser fields \cite{Lu2009_JPhysB, Feng2015_ChinJCP, Jin2016_ChinPhysB, Wang2012_AIP, Orlando2009_JPhysB}, plasmonic fields \cite{Wang2017_PRA, Yavuz2016_PRA, Feng2015_PRA, Yuan2017_JOSAB} on the harmonic cutoff and the intensity of the emitted harmonics is reported in the past. Moreover, some studies show that the phase of emitted harmonics is greatly influenced by the harmonic emission time and the carrier-envelope phase of the driving laser pulse \cite{De1998_PRL, Schafer2004_PRL}. The fundamental motivation behind the enhancement of the harmonics (in both the cutoff and the intensity) is the generation of an intense \textit{single attosecond pulse} instead of the attosecond pulse train \cite{Yuan2013_PRL, Khokhlova2016_PRA, Dashcasan2014_OptComm, Zeng2008_JPhysB, Goulielmakis2008_Science, Zhao2012_OptLett, Lopez-Martens2005_PRL}.  

\begin{figure}[b]
\centering\includegraphics[width=0.65\columnwidth]{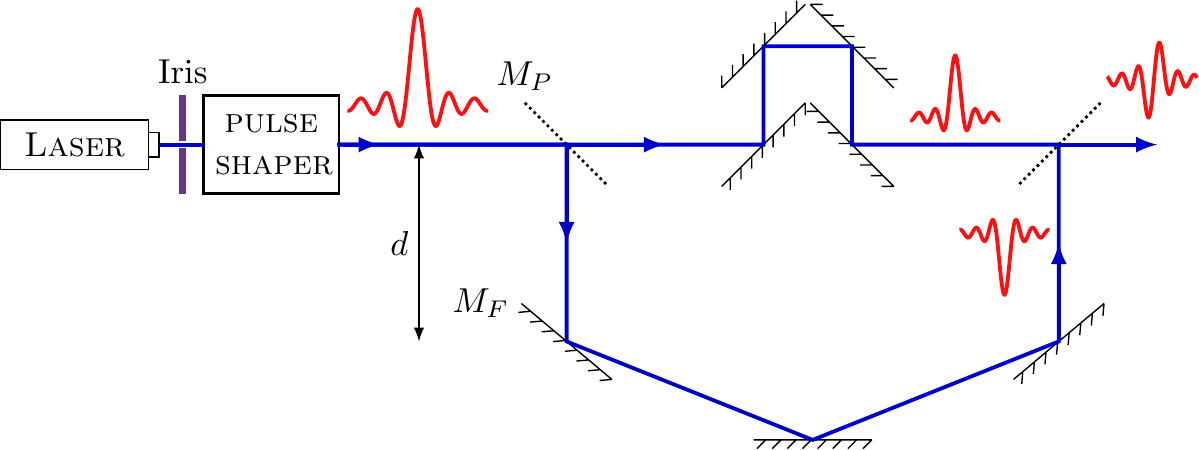}
\caption{Schematic diagram representing the proposed experimental setup. In this setup $M_P$ and $M_F$ are respectively mirrors with 50\% (beam splitter) and 100\% reflectivity. The configuration of an odd number of mirrors in the bottom arm is used to rotate the polarization of the pulse by 180$^\circ$. In the top arm, even number of mirrors are used to induce the desired path difference such that the pulse delay $\tau = 0$ would correspond to the out of  phase addition of both pulses after the second beam splitter.  }
\label{geo}
\end{figure}

The harmonics at the cutoff of the HHG spectrum are emitted in a short time with a relatively constant phase. The superposition of these harmonics can lead to the generation of an isolated attosecond pulse \cite{Hentschel2001_Nature, Kienberger2002_Science, Christov1997_PRL, Goulielmakis2004_Science}. On the other hand, the harmonics lie in the plateau region are generated by two primary electron paths (the so-called short and long paths) having two clearly distinguished ionization and recombination times. The superposition of these plateau harmonics leads to a short light burst of sub-femtosecond or attosecond duration separated by twice the laser frequency and for a multi-cycle driving pulse, the superposition of plateau harmonics is observed to produce an attosecond pulse train \cite{Agostini2004_RepProgPhys, Paul2001_Science}.

The typical shape of the sinc function makes it very interesting from the perspective of the generation of higher-order harmonics. The sinc pulse has a flattop spectral distribution, and also has a single relatively extreme field amplitude which can assist in accelerating the electron to achieve higher kinetic energies. If the electron recombines with the parent ion, then the excess energy will be emitted in the form of energetic photons. The pulse shaping techniques such as \textit{optical frequency combs} are around the corner for approximately three decades \cite{Cundiff2003_RMP,Krauss2009}. However, after the advent of the \textit{optical arbitrary waveform generation}, the pulse shaping is mainly achieved by the train of identical optical pulses produced by the mode-locked lasers \cite{Cundiff2010_NATP}. These mode-locked lasers might have the pulse duration of a few femtoseconds with a repetition rate of a few gigahertz. In order to produce an arbitrary optical waveform, one needs a pulse shaper which can be updated for each pulse. This facet of pulse shaping is actively worked upon by various researchers in the laser fraternity. We believe that the sinc pulses can be generated by the state of the art pulse shaping techniques such as deformable mirrors, a spatial light modulator, an acousto-optic modulator, and many more. By directly synthesizing rectangular shaped and phase-locked frequency comb, researchers have reported the generation of the sinc shaped pulses of exceptional quality \cite{Soto2013_ncomm}. The contemporary technological advances further increase the feasibility aspects of high power sinc shaped pulses in the near future.   

In this work, we discuss a simple setup using the sinc laser pulse(s), which promises a \textit{highly tunable} harmonic cutoff in similar intensity ranges. This, in turn, translates to the realization of the tunable radiation source having the photon energies ranging from XUV to soft-Xrays of the electromagnetic spectrum. The theoretical and simulation aspects of the work are discussed in section \ref{theory}, followed by the results and discussion in section \ref{result}. The self-contained theoretical formulation using strong-field-approximation along with the numerical implementation is discussed in section \ref{SFA} and finally concluding remarks are made in section \ref{con}.  

\section{Theory and Simulation Details}
\label{theory}
 The schematic diagram representing the setup is presented in figure  \ref{geo}. At present, there are no oscillators that can directly generate a sinc-shaped laser pulse. In view of this, a pulse shaper is introduced which transforms the incoming standard Gaussian laser into the sinc-pulse.  The Iris is placed in front of the pulse shaper to regulate the pulse energy. The shaped output pulse is then passed through the beam splitter.  One of the pulses acquires a phase of $180^\circ$ due to an odd number of reflections from mirrors, and then interferes with the other pulse \cite{Neyra2018_PRA}. The distance $d$, as shown in the figure (mirrors assembly in the bottom arm), can easily be tweaked on an optical bench, which in turn induces a path difference between the two pulses and that results in a delay between the pulses. This indicates that the delay parameter can be tuned easily. We will see that the delay between two pulses indeed causes a shift in the harmonic cutoffs. The temporal profile of the electric field of the synthesized pulse is then represented as:
\be
E(t) = E_0(\tau) \left[ \frac{\sin[\omega_0 (t-t_0-\tau)]}{\omega_0 (t-t_0-\tau) } -  \frac{\sin[\omega_0 (t-t_0)]}{\omega_0 (t-t_0) } \right], 
\label{field}
\ee
where the laser frequency $\omega_0 = 0.057$ a.u., $\tau$ is the delay between the pulses,  $E_0(\tau)$ is a delay-dependent field amplitude, and $t_0$ introduces some constant phase. Note that, $\tau = 0$ corresponds to the out of phase addition of the pulses which results in $E(t) = 0$ [refer figure  \ref{geo}]. Irrespective of the delay parameter $\tau$, the pulse energy $\Big(\sim \int |E(t)|^2 dt \Big)$ can be fixed at some constant value by tweaking the amplitude $E_0(\tau)$ using the Iris \cite{Nefedova2018_APL}. It should also be noted that $\tau$ can easily be tuned by varying the distance $d$ on an optical bench. Throughout the manuscript, we will be using the atomic units unless otherwise stated: this implies $e = \hbar = m_e = 1$.
  
We study the interaction of the synthesized laser pulse with the He-atom by numerically solving the one-dimensional time-dependent Schr\"{o}dinger equation (TDSE) based on the single-active electron (SAE) approximation \cite{Awasthi2008_PRA}. The TDSE in the length gauge is written as \cite{Chacon2015_PRA}:
 
\be
 i\frac{\partial \psi(x,t)}{\partial t}  = \Big[-\frac{1}{2}\frac{\partial^2}{\partial x^2}  + V(x)- x E(t) \Big] \psi(x,t), \label{tdse}\ee
where $E(t)$ is the laser field as given by equation \ref{field} and 
\be V(x) = - \frac{1}{\sqrt{x^2 + \xi}} \label{vsoft}\ee 
is the ionic soft-core potential where the constant $\xi$ is dependent on the ionization potential of the atom under study.  For He-atom $\xi = 0.484$ is considered such that the ground state energy (ionization potential) is found to  be $\sim 0.9$ a.u. ($\sim24.6$ eV), which is close to the experimental value of the ionization potential of Helium.  The general solution of TDSE is obtained by employing the time evolution operator $U(t_0 +\Delta t, t_0)$ on the initial state wavefunction of the electron $\psi_0(x,t_0)$,
\begin{equation}
\psi(x,t_0 + \Delta t) = U(t_0 + \Delta t, t_0) \psi_0(x,t_0),
\end{equation} 
The TDSE is solved numerically by adopting the split-operator method \cite{Feit1982_JCP} in which the evolution operator factored as a product of the kinetic and potential energy propagators, i.e.,
\be 
U(t_0 + \Delta t, t_0) = \text{e}^{-i p^2 \Delta t/4} \text{e}^{-i V_{\rm eff}(t_0 + \Delta t/2) \Delta t} \text{e}^{-i p^2 \Delta t/4},
\label{so:lg}
\ee 
where $V_{\rm eff} (t) = V(x) - xE(t)$ is the effective potential in the length gauge.  

 The initial ground state is calculated by the imaginary-time propagation method \cite{Bader2013_JChemPhys}. Starting from this initial state, the time-dependent wavefunction is obtained by solving the TDSE numerically. The time-dependent dipole acceleration $\ddot{d}(t)$ is evaluated following the Ehrenfest theorem as \cite{sandPRL_1999}:
\begin{equation}
\ddot{d}(t) = - \langle \psi(x,t) \Big|\frac{\partial V(x)}{\partial x}+ E(t) \Big| \psi(x,t) \rangle.
\label{dipacc}
\end{equation} 
The harmonic spectrum is then finally be obtained by the Fourier transformation of $\ddot{d}(t)$, i.e.,
\begin{equation}
S(\omega) = \Big|\frac{1}{\sqrt{2\pi}} \int \ddot{d}(t)\exp{[-i\omega t]} dt \Big|^2.
 \label{eq:hhg}
\end{equation}
In our calculation, the simulation domain is confined in a finite space of $4000$ a.u., where the grid spacing $\delta x = 0.05$ a.u. The simulation time step is decided from the relation $\delta t \sim (\delta x)^2/2$. In order to avoid the reflection of the electron wave packet from the boundaries, an absorbing boundary of thickness $200$ a.u. was placed at $x=\pm 1800$ a.u. The absorption is implemented using the exterior complex scaling method \cite{He2007_PRA}. The convergence of the calculation is checked by varying the grid parameters.

\begin{figure}[t]
\centering\includegraphics[width=0.65\columnwidth]{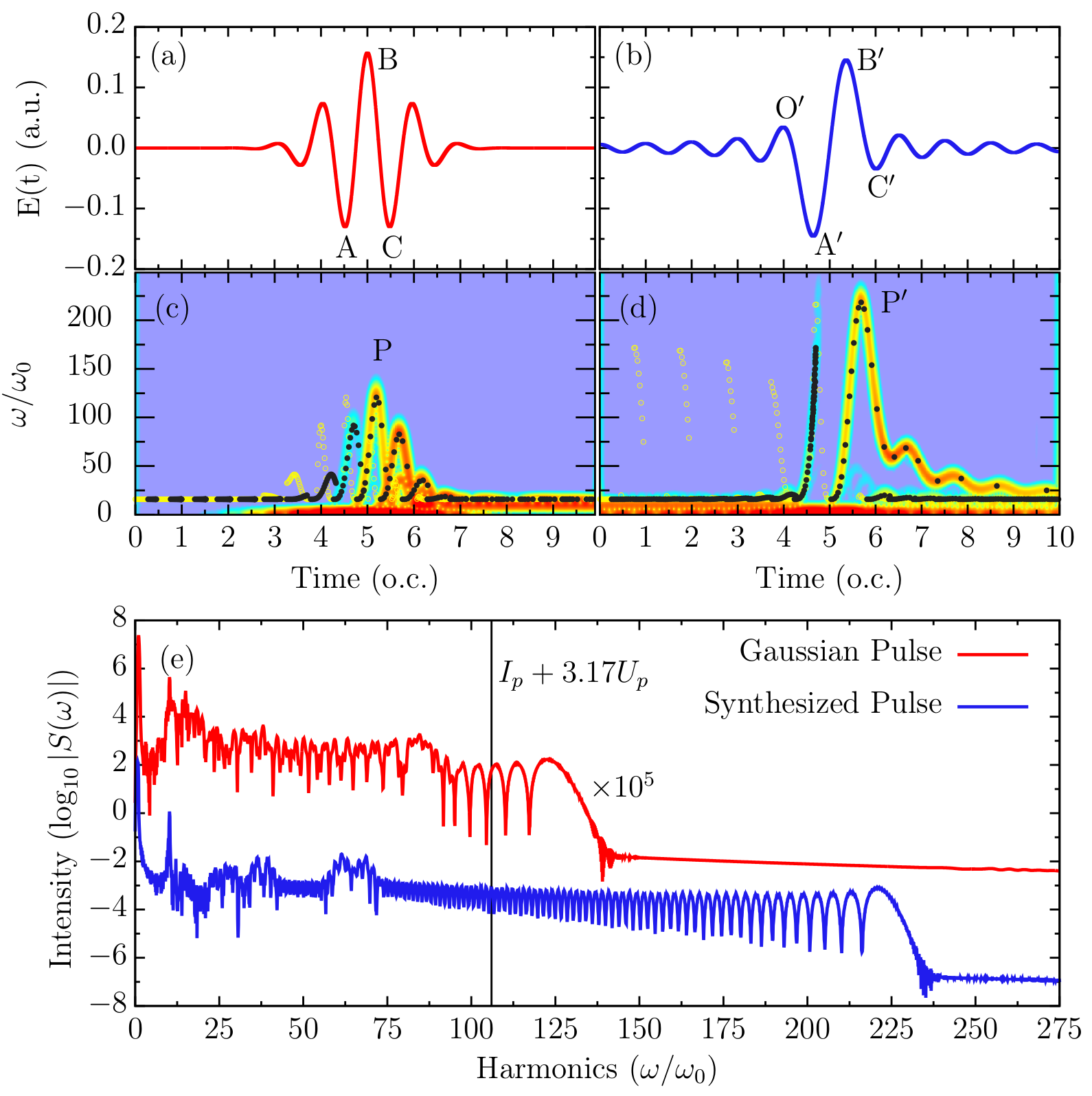}
\caption{The electric fields of the 800 nm / 5 fs Gaussian-envelope pulse (a) and synthesized pulse with parameter $\tau$ = 0.71T (b) are presented. We have presented the corresponding classical ionizing and returning energy maps along with the time-frequency distributions of the HHG spectra in (c) and (d) respectively for Gaussian and synthesized pulse. However, the harmonic spectra for both cases are compared in (e). For the purpose of clarity, the harmonic intensity of the Gaussian pulse is multiplied by a factor of $10^5$. The cutoff as predicted by the 3 steps model is also shown as a vertical line
($106\omega_0$) in (e).}   
\label{GaussCompare}
\end{figure}

The attosecond pulse is obtained by superposing several harmonics as \cite{Jooya2016_PL}:
\begin{equation}
I(t) =\Big|\sum_{q} a_q \exp{[iq\omega t]}\Big|^2,
\label{ASP}
\end{equation}
where $q$ is the harmonic order and $a_q$ represents the inverse Fourier transformation given as:
\be a_q = \frac{1}{\sqrt{2\pi}}\int \ddot{d}(t)\exp{[-iq\omega t]} dt. \ee 
The time-frequency analysis has been done to get insight of the quantum recollision processes. The wavelet transform has been performed using the standard Gabor time-frequency analysis \cite{Koushki2016_comptc, Zhong2016_PLA}.
 
\section{Results and Discussions}
\label{result}

In the following, we compare the harmonic spectrum by a Gaussian laser pulse and the synthesized pulse given by equation \ref{field}, followed by the effect of the delay $\tau$ on the harmonic cutoffs, and finally its impact on the generation of the optimal attosecond pulses. 

\subsection{Comparison with Gaussian Pulse}

Initially, in order to show the preeminence of our synthesized pulse for the generation of higher-order harmonics, we have compared the HHG spectra generated by the synthesized laser field defined in equation\ref{field} and the $800$ nm/$5$ fs Gaussian-enveloped laser pulse. The temporal profile of the Gaussian laser pulse is given as: 
\be E_g(t) = E_{0g} \exp\left[ - 4 \ln 2  \frac{t^2}{\tau_g^2}\right]\cos(\omega_0 t), \ee 
where $E_{0g}$ is the amplitude of the laser field, and $\omega_0$ (= 0.057 a.u.) is the laser frequency same as used in the synthesized pulse, and $\tau_g = 5$ fs is considered. It should be noted that the pulse energy of both the pulses is fixed at the value $\sim 1.91$ a.u. Thus the peak field amplitude for the Gaussian pulse is estimated to be $E_{0g} \sim 0.1567$ a.u. However, the field amplitude of the synthesized pulse is considered to be $E_0 \sim 0.1453 $ a.u.

\begin{figure}[t]
\includegraphics[width=1\columnwidth]{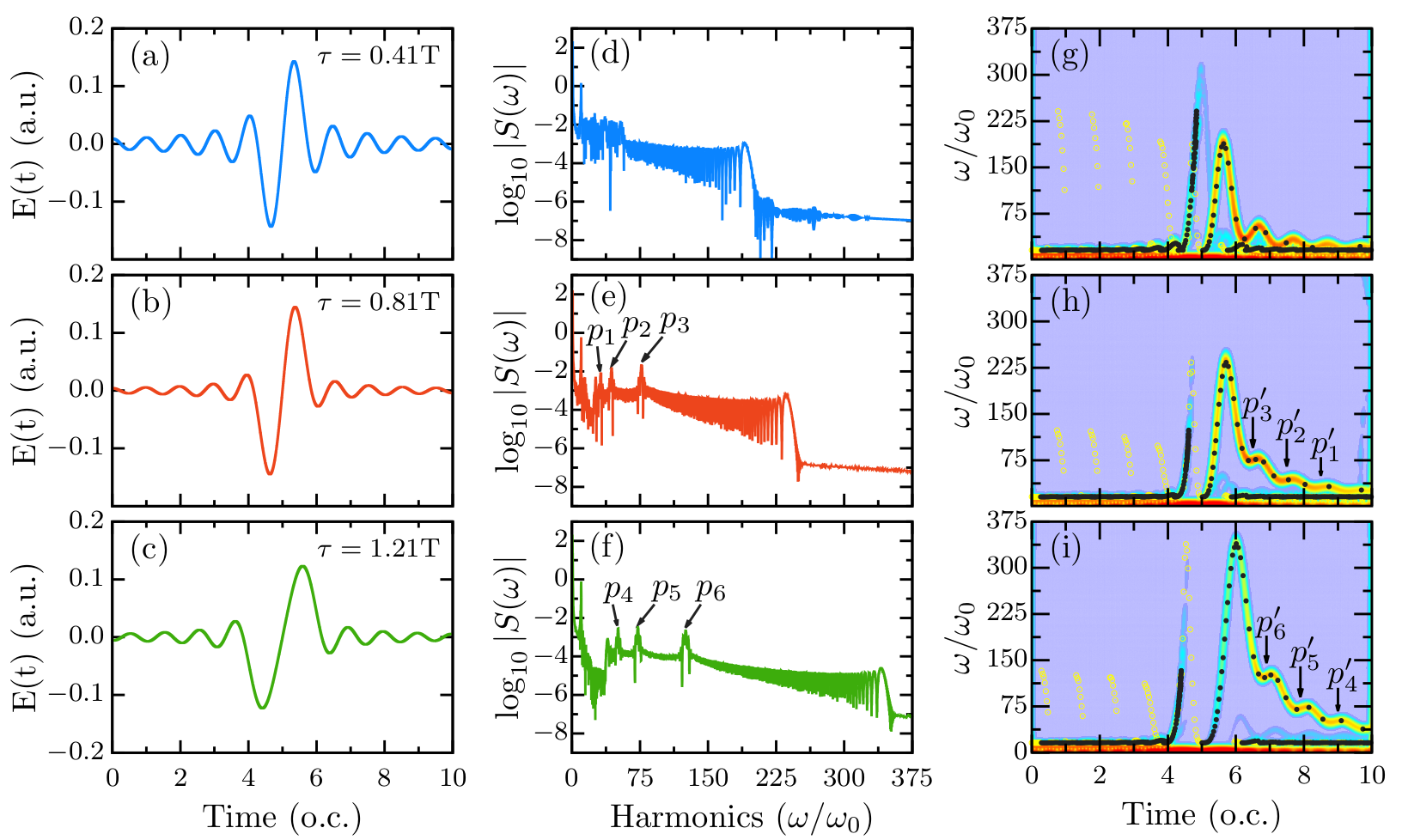}
\caption{Temporal profile of the electric field with delay parameter $\tau = 0.41$T (a), $0.81$T (b), and $1.21$T (c). Corresponding harmonic spectra are respectively presented in (d), (e) and (f). Time frequency analysis of the electron quantum path along with the classical ionization and recombination time are shown in (g), (h) and (i) respectively. }
\label{DelayCompare}
\end{figure}

The temporal electric field profile of the Gaussian-envelope and the synthesized pulse is presented in figure  \ref{GaussCompare}(a) and \ref{GaussCompare}(b), respectively. For the synthesized pulse, the delay parameter $\tau$ is chosen to be $0.71 $T, where $T\ (= 2\pi/\omega_0)$ is the time corresponding to one optical cycle (o.c.). Compared to the case of the Gaussian pulse, we can see that the laser cycle of the synthesized pulse is expanded. The electron ionized around the negative maximum of the synthesized laser field can gain higher energies in the relatively longer acceleration process. As a consequence, the harmonics with larger cutoff energy would then be achieved when it will recombine with the parent ion. The comparison among the HHG power spectra for the Gaussian and the synthesized pulse is presented in figure  \ref{GaussCompare}(e). The differences in harmonic intensities among the two curves in the plateau region are actually small, so for the purpose of clarity we have shifted the harmonic signal for the Gaussian pulse along the $y$-axis. These harmonic spectra have the characteristic structure of a typical HHG spectrum, i.e., irregular behavior towards the lower harmonics region, then gradually it becomes regular within the plateau region, and finally a sudden drop in the harmonic intensity at the {\it cutoff}. The results show that the harmonic cutoff is impressively extended from $140\omega_0$ for the Gaussian field to $235\omega_0$ for the synthesized field. The harmonic structure in the supercontinuum region of the synthesized field is less modulated than the case of Gaussian field. This condition is favorable for the generation of isolated attosecond pulse. The cutoff rule \cite{Krause1992_PRL} $I_p + 3.17U_p$ (where $U_p = E^2/4\omega_0^2$ is the quiver energy, and $I_p = 0.904$ a.u. is the ionization potential for the He-atom) for the maximum possible harmonic photon energy predicts the cutoff at $106\omega_0$ ($164$ eV), while the harmonic cutoff for the case of synthesized pulse is observed at $235\omega_0$ ($364$ eV). This inability of three steps model to predict the harmonic cutoff is due to the classical consideration of the fact that the laser pulse intensity should be constant during the quiver motion of electron \cite{Christov1997_PRL, Zhou1996_PRL}. Since the laser intensity in our case changes significantly in one laser cycle, we can not expect that the celebrated three steps model would predict the harmonic cutoff correctly. 

Both the classical theory \cite{Corkum1993_PRL} and quantum time-frequency analysis are adopted to have a deeper understanding of the generation of the higher-order harmonics. In figures \ref{GaussCompare}(c) and \ref{GaussCompare}(d), we show the classical electron trajectories along with the time-frequency distribution of the HHG spectra for the above two cases. In the case of the synthesized field, the electron trajectory map shows two paths with different emission times (solid black circles) contributing to each harmonic in HHG spectra [refer figure \ref{GaussCompare}(d)]. The two branches of the emission time trajectory with positive and negative slopes are corresponding to the short and long paths, respectively. Likewise, in the time-frequency profile, there are two quantum paths contributing to each of the harmonic within half of the optical cycle of the synthesized laser field. It should be noted that the maximum harmonic order corresponding to the Gaussian pulse, marked by the point $P$ in figure  \ref{GaussCompare}(c), is approximately equal to $140 \omega_0$; whereas, the maximum harmonic order corresponding to the synthesized pulse, marked as $P^\prime$ in figure  \ref{GaussCompare}(d), is observed at the value $235 \omega_0$. Even though the field energy content of both the driving pulses are the same, but the order of the cutoff harmonic corresponding to the synthesized pulse is about $95 \omega_0$ (or $147$ eV) more than the same corresponding to the Gaussian pulse. The harmonic spectrum for these cases is illustrated in figure  \ref{GaussCompare}(e), which complement the results presented using the time-frequency analysis.

\begin{figure}[b]
\centering\includegraphics[width=0.65\columnwidth]{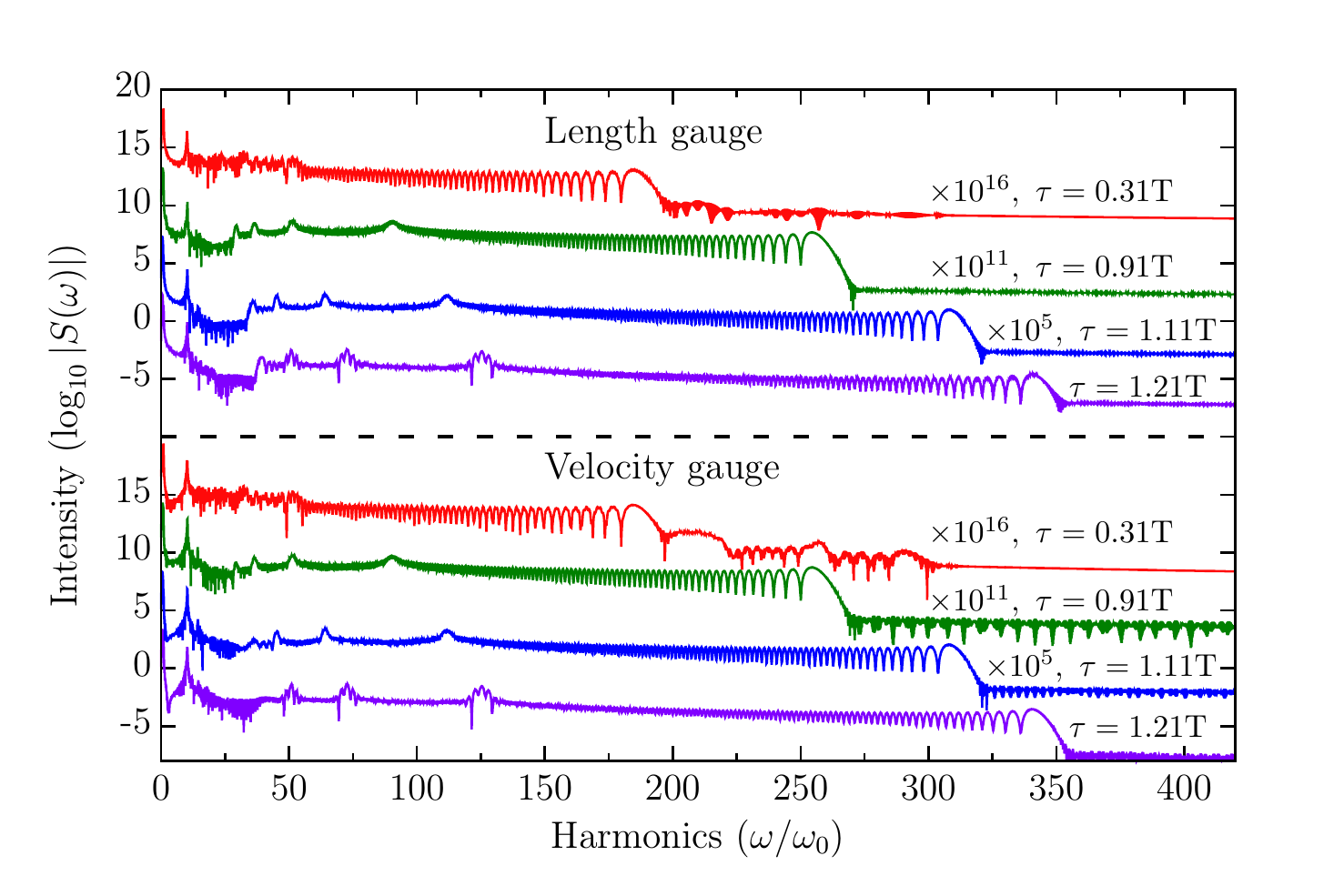}
\caption{The HHG spectra of synthesized pulse for different delay cases are presented  using length (top) and velocity gauge (bottom) while solving the TDSE. For the purpose of clarity, the harmonic intensities of $\tau = $0.31T, 0.91T, and 1.11T pulses are multiplied by factors of $10^{16}$, $10^{11}$ and $10^5$, respectively. }
\label{lg_vg}
\end{figure}

Based on the classical trajectory maps, we can state that the peak $P$ is originated by the $ABC$ process marked in figure  \ref{GaussCompare}(a). The electron gets ionized by tunneling when the magnitude of the laser field amplitude reaches its maxima around the point $A$, then it gets accelerated by the next maxima at $B$ which results in gaining high kinetic energy, and finally the electron recombines with the parent ion core in between the peak at $B$ and the dip at $C$, which results in the emission of harmonic photons up to the order of $135$. The maximum energy at the peak $P$ is determined by the kinetic energy gained by the electron during the acceleration process. On the contrary, in the case of the synthesized field [refer figure  \ref{GaussCompare}(b)], the laser cycle is expanded, and therefore the electrons get ionized around $A^\prime$. The free electron then accelerates for a relatively long time till the peak $B^\prime$. Then it returns to the core ion with higher kinetic energy in between $B^\prime$ and $C^\prime$ and recombines with the ion. The recombination process leads to the emission of harmonics up to the order of $235$. 

\begin{table}[b]
\caption{\label{table1}Summary of harmonic cutoffs ($\omega_c$) for different delay parameters ($\tau$) are presented along with their respective intensities ($\log_{10}|S(\omega_c)|$).  The  enhanced harmonics ($\omega_{en}$) are also summarized in last column and their respective intensities in multiple ($\eta$) of their cutoff intensity are mentioned in the bracket. All the harmonics are presented in units of the fundamental frequency $\omega_0$.  }
\footnotesize
\begin{indented}
\item[]\begin{tabular}{@{}llll}
\br
$\tau$ [T] &  $\omega_c$  & $\log_{10} |S(\omega_c)| $  &  $\quad\quad\quad\omega_{en}(\eta)$\\ 
\mr
   0.81 &  248  &  -3.18  &  26(8), 32(14),  44(24), 77(35)   \\
   0.91 &  271  &  -3.33  & 29(4), 36(7), 51(11), 90(10)     \\
   1.01 &  291  &  -3.58  & 32 (2), 40(3), 57(5), 100(5)     \\
   1.11 &  320  &  -4.00  & 36(5), 45(14), 64(21), 112(17)   \\  
   1.21 &  350  &  -4.55  & 39(33), 50(112), 72(132), 125(85) \\  
\br
\end{tabular}
\end{indented}
\end{table}
\normalsize

\subsection{Effect of Delay on HHG and Scaling law}

We now show how the synthesized pulse waveform resulting from the time delay ($\tau$) between two-component sinc pulses affects the resulting HHG spectra. The temporal profiles of the synthesized pulse for different time delays $\tau$ = 0.41T, 0.81T, and 1.21T are respectively presented in figures \ref{DelayCompare}(a) - \ref{DelayCompare}(c). The effect of the increasing delay on field strength can be seen as an extension in the laser cycle. The respective harmonic spectra [refer equation \ref{eq:hhg}] for these delay parameters are presented in figures  \ref{DelayCompare}(d)-\ref{DelayCompare}(f). The harmonic cutoff energies for these delays are observed respectively at $202 \omega_0$, $245 \omega_0$, and $350\omega_0$. Furthermore, one important feature is observed in the harmonic spectrum presented in figures \ref{DelayCompare}(e) and \ref{DelayCompare}(f) is the enhancement in HHG yield of harmonics around the peaks $p_1, p_2, p_3$ marked in figure  \ref{DelayCompare}(e) and the peaks marked as $p_4, p_5, p_6$ in figure  \ref{DelayCompare}(f). The underlying mechanism behind the extension of the HHG cutoff energy and the selective harmonic yield can be understood from the classical trajectories and the quantum time-frequency analysis of the quantum paths of the electron, as shown in figures \ref{DelayCompare}(g)-\ref{DelayCompare}(i). Here, the classical trajectory analysis shows how the effect of the time dependence of electron kinetic energy on the ionization (yellow circles) and also on the recombination (solid black circles) times.

The classical electron trajectories are consistent with the results of the time-frequency analysis of the electron quantum paths. The increasing delay time ($\tau$) results in the broadening of both negative and positive cycle of the synthesized laser field. This eventually modifies the ionization and recombination time, and also the kinetic energy of returning electron. The increased electron kinetic energy will cause an enhancement in the emitted harmonic photon energy as shown in figures \ref{DelayCompare}(g)-\ref{DelayCompare}(i). Furthermore, relatively higher intensities of harmonics around the peaks at $p_1, p_2$ and $p_3$ in figure  \ref{DelayCompare}(e) can be understood by the respective time-frequency analysis as shown in figure  \ref{DelayCompare}(h), where the corresponding plateaus are marked as $p_1^\prime, p_2^\prime$ and $p_3^\prime$. As can be seen that the harmonic emissions for the same harmonics ($p_1, p_2 $ and $p_3$) are taking place for relatively longer time. Similarly, in figure  \ref{DelayCompare}(i), the harmonics at plateaus $p_4^\prime, p_5^\prime$ and $p_6^\prime$ are getting emitted for relatively longer time, giving higher yield of those selected harmonics as shown around the peaks at $p_4, p_5$ and $p_6$ in figure \ref{DelayCompare}(f).

 In order to validate the numerical accuracy of the results obtained, we have also calculated the harmonic spectrum using the  velocity gauge while solving the TDSE. For the case of the spatially homogeneous fields, the time evolution operator in equation \ref{so:lg} will have the following equivalent form in the velocity gauge \cite{Chacon2015_PRA}: 
\be 
U(t_0 + \Delta t, t_0) = \text{e}^{-i (p + A(t_0 + \Delta t/2))^2 \Delta t/4}  \text{e}^{-i V(x) \Delta t}\\
\times \text{e}^{-i (p + A(t_0 + \Delta t/2))^2 \Delta t/4},
\label{so:vg} 
\ee 
where $A(t) = - \int_0^t E(t') dt'$ is the vector potential of the laser field. Further, the dipole acceleration expectation value and eventually the spectral intensity of harmonic emission can be calculated by following the similar prescription as mentioned for the case of length gauge in Sec. \ref{theory}. 

 In figure  \ref{lg_vg}, we present the harmonic spectrum using the length (top panel) and velocity gauge (bottom panel) while numerically solving the TDSE for different delay parameters. The HHG spectra is obtained for $\tau$ = 0.31T, 0.91T, 1.11T, and 1.21T using both the gauges with same the laser parameters.  It can be inferred from figure  \ref{lg_vg} that both the length and velocity gauge give the same results for the harmonic spectrum, which corroborates the convergence of the numerical techniques used to describe the HHG process. 

\begin{figure}[t]
\centering\includegraphics[width=0.65\columnwidth]{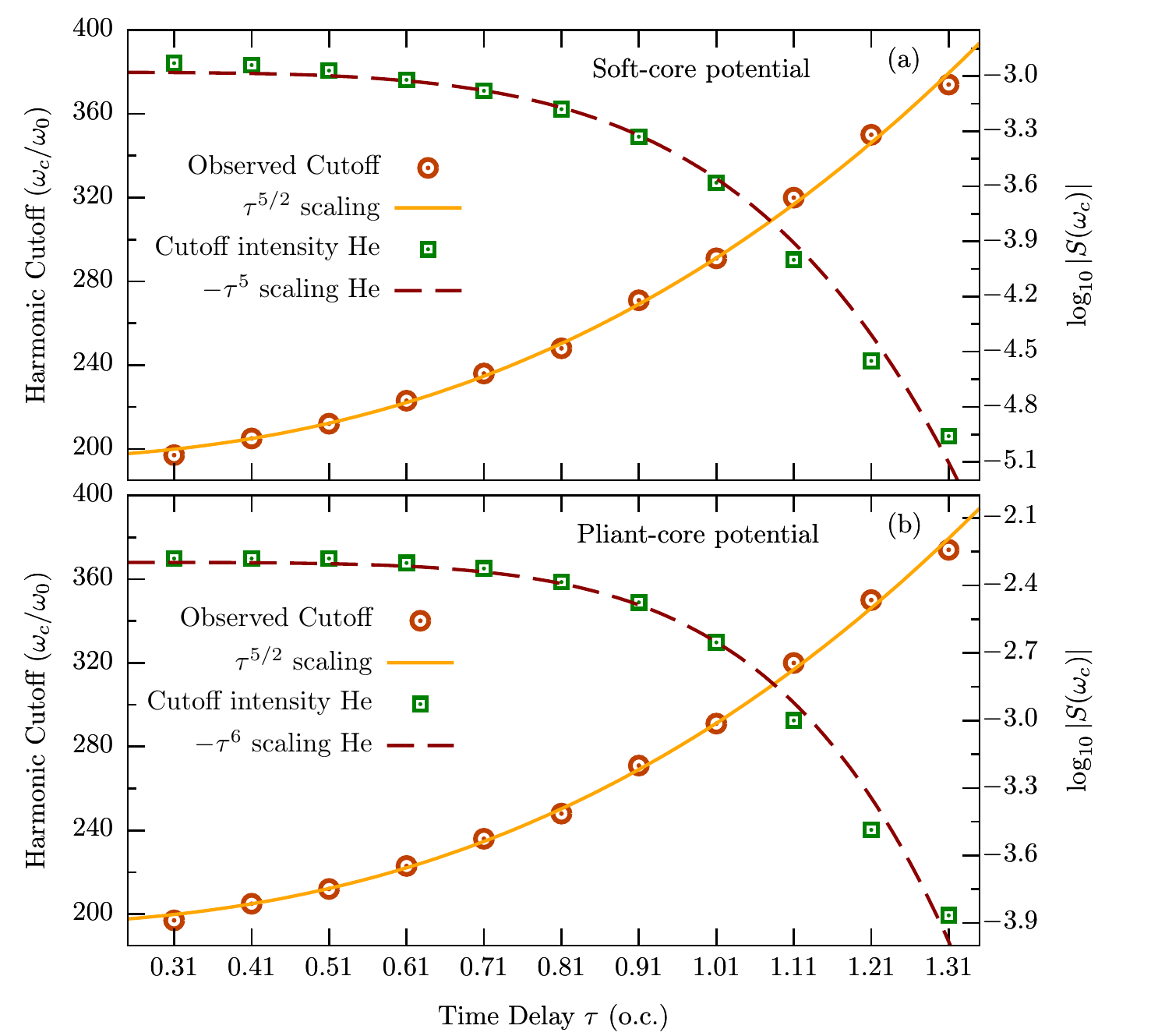}
\caption{ The observed harmonic cutoffs by varying the delay ($\tau$) between the two pulses [refer figure  \ref{geo}]
is presented using (a) {\it soft-core} [equation \ref{vsoft}] and (b) {\it pliant-core} potentials [equation \ref{vpc}]. The intensities of respective harmonic cutoffs are also illustrated for both the potentials. The cutoff and intensity scalings are also depicted for all sets of observed parameters [refer text for more
details]. Here, the peak field amplitude $E_0(\tau)$ is varied with $\tau$ in order to have the constant
pulse energy ($\propto \int |E(t)|^2 dt$).  Field amplitude and the pulse energy are normalized with respect to values associated with the delay parameter $\tau = 0.71$T. The peak field amplitude and the pulse energy for $\tau = 0.71$T are respectively $\sim 0.1453$ a.u. and $\sim 1.91$ a.u. }
\label{Scale}
\end{figure}

 Moreover, figure  \ref{lg_vg} presents the harmonic spectrum for different delay parameters. We can notice that, not only the harmonic cutoff changes with the delay parameter $\tau$, but some harmonic components are also observed to be enhanced as compared to the neighboring harmonics in the plateau region. The position of these enhanced harmonics is found to be changing for different time delays. The intensity of these harmonics is about one order of magnitude higher than the surrounding harmonics. The mechanism behind these selective enhancement has already been discussed above. These enhanced harmonics can be used further as a monochromatic source for many important applications, which include the seeding of an XUV free-electron laser or laser-plasma amplifiers \cite{Lambert2008_Nature}. In Table \ref{table1}, we have presented some typical harmonic cutoffs and their intensities for different values of the delay $\tau$. Here we have also presented some enhanced harmonics and their intensities as a multiple of the intensity observed at the harmonic cutoffs.

In figure  \ref{Scale}(a), the observed harmonic cutoff $\omega_c/\omega_0$ for different delay parameters ($\tau$) is presented  for using the soft-core potential [equation \ref{vsoft}].  Based on these observations, we infer that the HHG cutoff in the current setup scales as $\sim\tau^{5/2}$ for a fixed driver pulse energy  irrespective of atomic species. The scaling law is found to be consistent upto some large time delay such as $1.31 $T.   Moreover, further increase in delay deforms the synthesized pulse by inducing local field maximum between the two central peaks of the field.  The intensities of the respective harmonic cutoff for different delay parameters are also illustrated in figure  \ref{Scale}(a). It is observed that the cutoff intensities ($\log_{10}|S(\omega_c)|$) follow a $\sim -\tau^5$ scaling. The decrease in HHG yield can be explained by taking into account the spreading of electronic wave-packet during propagation in the continuum. As we have mentioned earlier, increasing the delay in pulse synthesis would result in elongation in the laser cycle. Since the wave-packet propagation time in the continuum is proportional to the laser cycle, the wave-packet has more time to spread with increasing delay, thus scaling down the efficiency of the recombination process in harmonic emission.  

\begin{figure}[b]
\centering\includegraphics[width=0.65\columnwidth]{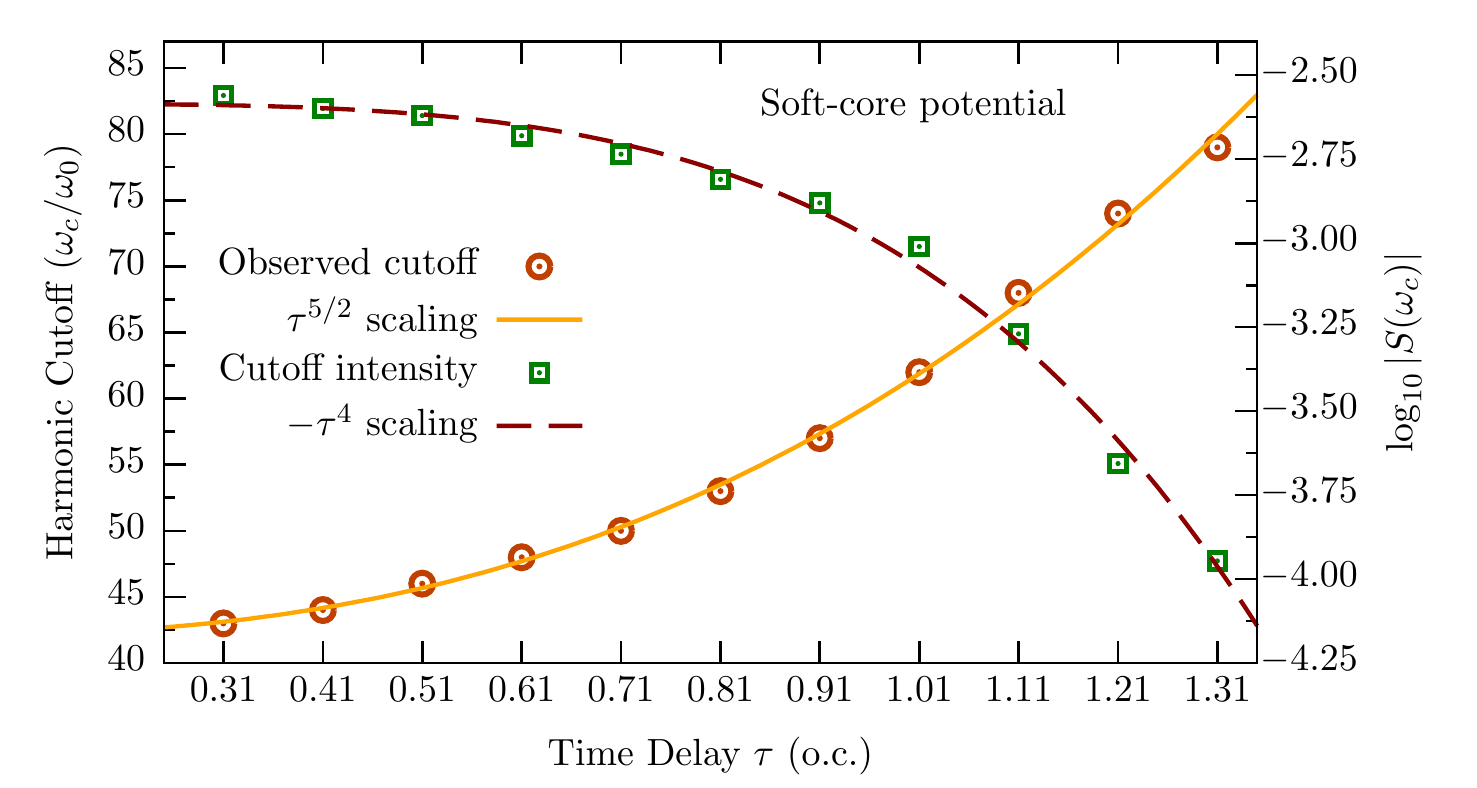}
\caption{Scaling of harmonic cutoff and the cutoff intensity with delay parameters for Hydrogen atom using soft-core potential [refer equation \ref{vsoft}] with $\xi = 2$. The laser intensity is considered to be $10^{14}$ W/cm$^2$ with the same field profile as used in figure  \ref{Scale}. The pulse energy is kept fixed. Field amplitude and the pulse energy are normalized with respect to values associated with the delay parameter $\tau = 0.71$T. The peak field amplitude and the pulse energy for $\tau = 0.71$T are respectively $\sim 0.065$ a.u. and $\sim 0.379$ a.u. } 
\label{Scale_H}
\end{figure}


 Furthermore, to analyze the validity of these scaling laws in the three-dimensional environment, we have used a 1D modified potential referred as {\it pliant-core} potential \cite{Silaev2010_PRA}. This modified potential ensures a good agreement with the 3D TDSE calculations as reported in \cite{Silaev2010_PRA} and has the following form:
\be V_\text{pc}(x) = -\frac{1}{( |x|^{3/2} + \beta )^{2/3}} \label{vpc}\ee 
where the parameter $\beta$ is used to avoid the Coulomb singularity. Also, the value of parameter $\beta$ varies according to the ionization potential of the atom under study. For He atom $\beta =  0.49$ is used \cite{Silaev2010_PRA}.  

 In figure  \ref{Scale}(b), we have presented the observed harmonic cutoff $\omega_c / \omega_0$ and their corresponding harmonic intensities obtained using the {\it pliant-core} potential [equation \ref{vpc}]. It is observed that the HHG cutoffs for different delay parameters are also scaled as $\sim \tau^{5/2}$, showing an excellent agreement with the results obtained in the case of the {\it soft-core} potential. The intensities of cutoff harmonics are also following a definite scaling of $\sim -\tau^6$ for   using {\it pliant-core} potential. As it was reported in \cite{Schiessl2007_PRL, Ishikawa2009_PRA}, the wavelength scaling of the harmonic intensity can vary significantly for different atoms.   In order to see the universality of the scaling law, we have also obtained the scaling laws for the Hydrogen atom (using $\xi = 2$ in equation \ref{vsoft}) with reduced laser intensity of $10^{14}$ W/cm$^2$, while keeping the field profile same as used in figure  \ref{Scale}.  The harmonic cutoff and intensity scaling for the Hydrogen atom are presented in figure  \ref{Scale_H}. It has been observed that the harmonic cutoff scales as $\sim\tau^{5/2}$ and the cutoff intensity scales as $\sim -\tau^{4}$.  


\begin{figure}[t]
\centering\includegraphics[width=0.65\columnwidth]{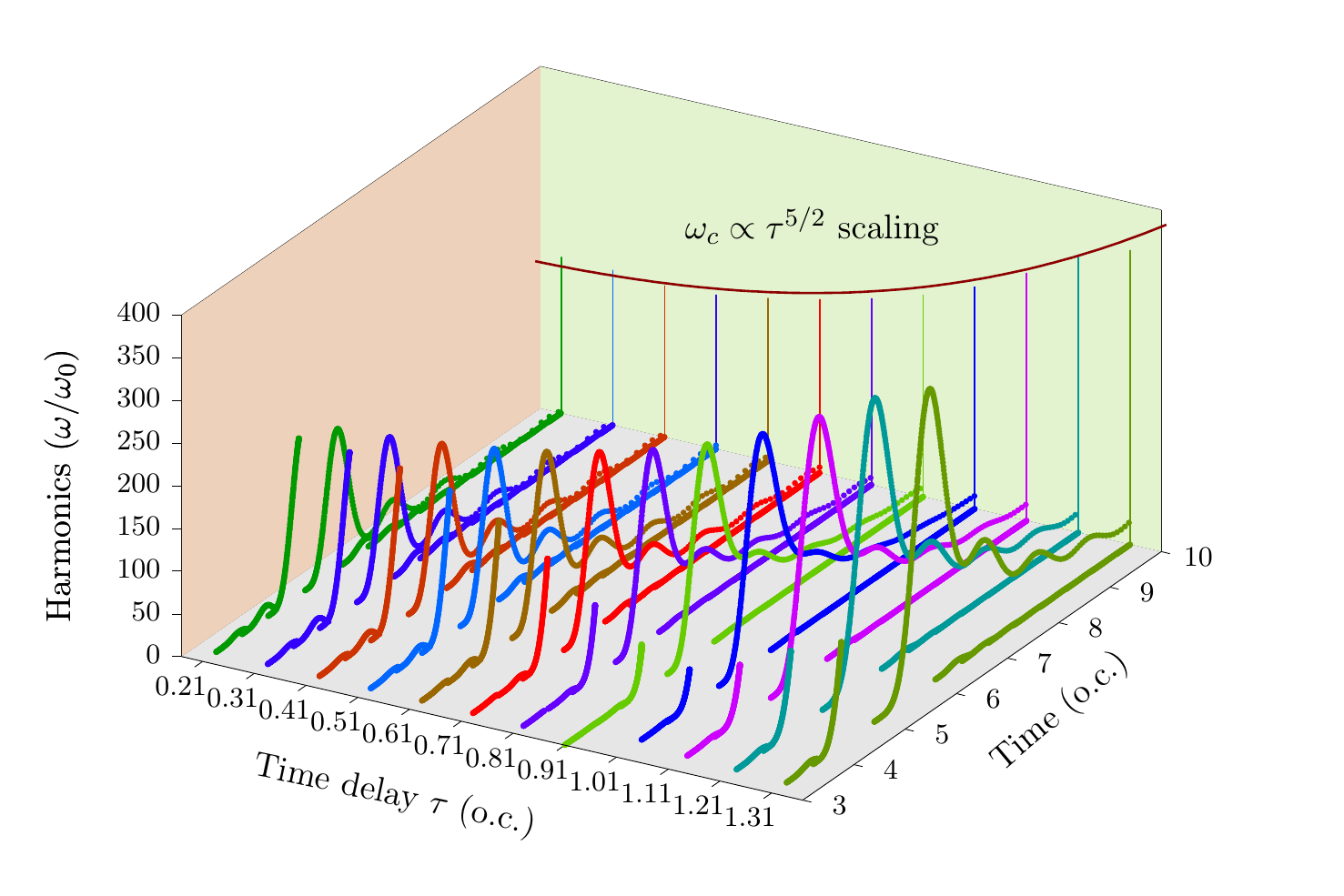}
\caption{Classical trajectories are presented for different delay parameters and again the $\omega_c\propto \tau^{5/2}$ scaling is also projected. All other laser parameters are same as presented in figure  \ref{Scale}. }
\label{clatraj}
\end{figure}

These scaling laws for harmonic cutoff and their respective intensities clearly demonstrate the utility of this setup. We not only predict the harmonic cutoff, but also its intensity in terms of the time delay between two oppositely polarized sinc pulses.  The above mentioned scaling law of the harmonic cutoff energy is also validated using the classical trajectory analysis and the corresponding returning electron trajectory maps are presented in figure  \ref{clatraj}. Even though there is no way to comment on the cutoff intensities from the classical analysis, but the scaling property of the maximum radiated photon energy can be predicted promisingly. It is clear from figure  \ref{clatraj} that, for smaller time delays (i.e., $\tau < 0.41 $T), the cutoff harmonics are emitted due to the ionization around the peak marked as $O^\prime$ in figure  \ref{GaussCompare}(b)] of the synthesized laser pulse. The electrons ionize around the peak at $O^\prime$ and later recombine between the dip at $A^\prime$ and the peak at $B^\prime$. These electrons accelerate for relatively longer times as compared to the electrons ionized around the dip at $A^\prime$ and recombined between the peak at $B^\prime$ and the dip at $C^\prime$. Therefore, the energy of the emitted photons is higher for the former case. 

Moreover, with the increasing delay, the width of the dip at $A^\prime$ and the peak at $B^\prime$ increases, and concurrently the amplitude of the peak at $O^\prime$ decreases [refer figure  \ref{DelayCompare}(a)-\ref{DelayCompare}(c)]. As a result, the cases for which $\tau$ is greater than $0.41$T, the quiver energy of the electrons ionized around the peak at $O^\prime$ decreases, while for the electrons ionize around the dip at $A^\prime$ would return with higher kinetic energies as observed in figure  \ref{clatraj}. The typical nature of the sinc pulse yields the previously mentioned scaling of the cutoff energy and its intensity in the harmonic spectrum. The delay parameter actually controls the interference between the two pulses of opposite polarization, which eventually translates into the modification of the electron quantum path as well as the recombination energy of the returning electron.   
 
\subsection{Attosecond pulse generation}

\begin{figure}[b]
\centering\includegraphics[width=0.65\columnwidth]{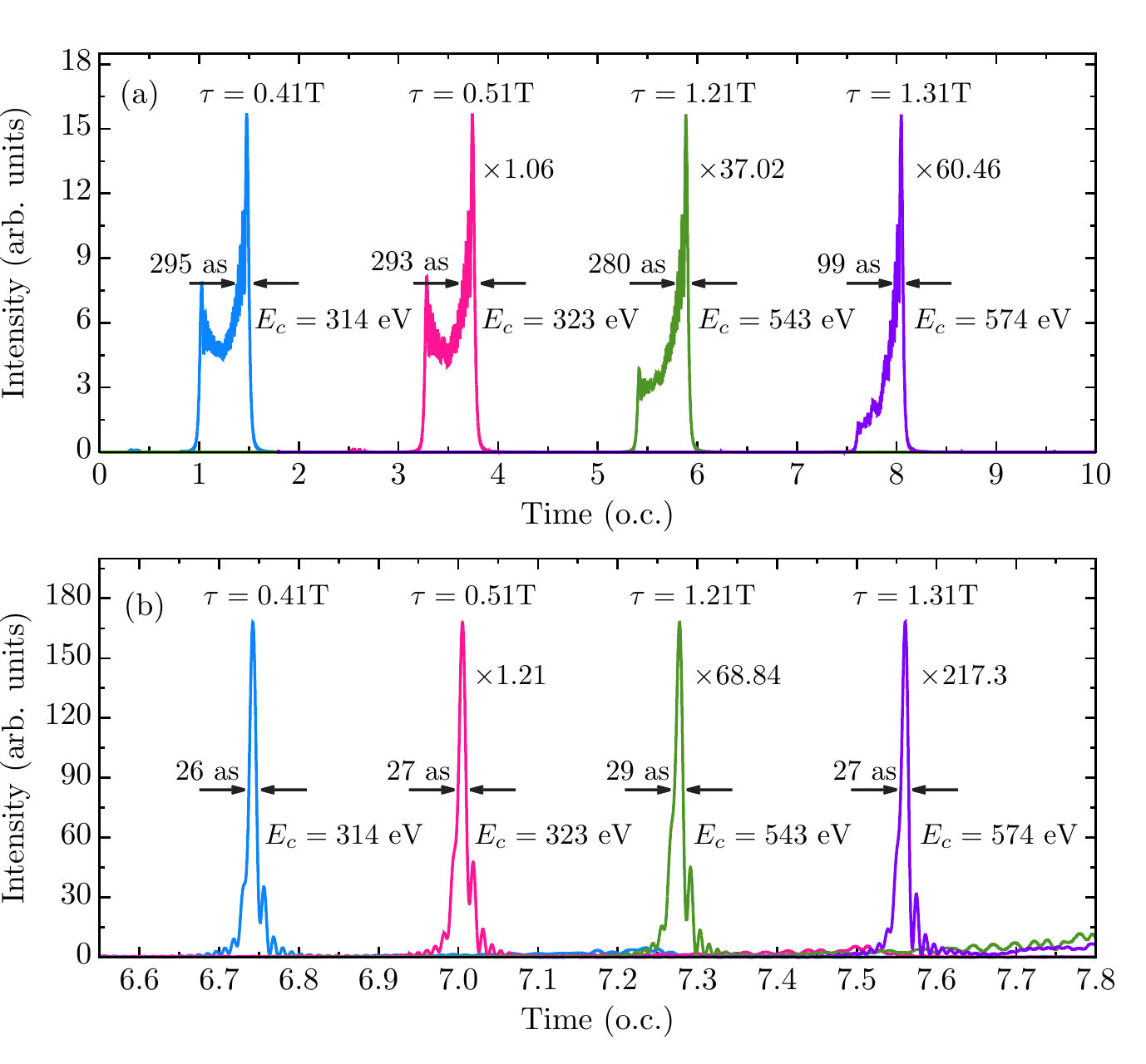}
\caption{The temporal profiles of the attosecond pulses generated by properly selecting the filtering window of $100$ harmonics before cutoff for time delays $\tau$ = 0.41T, 0.51T, 1.21T, and 1.31T : without (a) and with (b) phase compensation is presented. For the purpose of clarity, the ASPs are shifted in time scale and multiplied by some factors as shown.  The cutoff energy ($E_c$) is also mentioned for each ASP.}  
\label{AttosecondPulse}
\end{figure}

We now discuss the role of the higher-order harmonics in the generation of a single attosecond pulse. In figure  \ref{Scale}, we observe nice scaling behavior of the harmonic cutoff and their respective intensity in terms of the delay parameter $\tau$. We now study how to generate an attosecond pulse (ASP) by filtering several harmonics just before the cutoff but in the plateau region. The temporal profile of the ASP for a given delay parameter is obtained by superposing the contribution of the different harmonics and then performing the inverse Fourier transform [refer equation \ref{ASP}]. In figure  \ref{AttosecondPulse}(a), we present isolated attosecond pulses generated without any phase compensation, but just by superposing the last $100$ harmonics up to the cutoff of the harmonic spectrum corresponding to a bandwidth of $\sim 155$ eV for different delay parameters. The pulses have been emitted within the same optical cycle, but just for the sake of clear presentation, we have shifted the pulses along the time scale. The maximum harmonic photon energy for different ASPs corresponding to delay $\tau$ = 0.41T, 0.51T, 1.21T and 1.31T are 314 eV, 323 eV, 543 eV and 574 eV, respectively [refer figure  \ref{Scale}]. The generated ASPs become shorter with increasing values of the delay parameter. Here we show that, as we increase the delay from $\tau = 0.41$T to $\tau = 1.31$T, the pulse width of the generated ASPs decrease from $295$ as to $99$ as. We explain this behavior using figures \ref{DelayCompare}(g)-\ref{DelayCompare}(i). Here we see that as we increase the delay, the intensity of the short trajectory diminishes. This indicates that the contribution of the harmonics due to the shorter trajectories is very less in the HHG spectrum. Therefore, in the formation of attosecond pulses, the harmonics corresponding to the longer trajectories are present, and that results in the shorter ASP.  It should be noted that, by the nature of the synthesized pulse [refer equation(1)], harmonic emission takes place only for a single cycle of the driving pulse. Therefore, we will have at the most two ASPs instead of a pulse train upon the superposition of a large range of harmonics in the plateau region of the HHG spectrum. These paired ASPs are a potential tool in pump-probe measurements. As we already discussed, with increasing delay the contribution of only a single trajectory (i.e., long trajectory) is dominant in the HHG spectrum. Hence, the intensity of one of the ASP in the pair diminishes with increasing delay. The single ASP is more promising for applications in time-resolved spectroscopy. This method reduces the need to adopt various gating techniques to obtain the single ASP \cite{Huo2005_OptExp}. 

The harmonics in the HHG process are emitted over a range of time, which means the phase of each harmonic is arbitrary. As a result, the superposition of entire plateau harmonics can not generate the shortest attosecond pulse. However, the phase dispersion can be compensated by propagating the ASP through either a gas medium or a thin metal foil \cite{Lopez-Martens2005_PRL, Mairesse2003_Science}. In our simulations, the harmonic phase has been compensated by taking a constant phase difference between two consecutive harmonics. The resultant ASP after the phase compensation for different delay parameters are presented in figure  \ref{AttosecondPulse}(b). After the phase compensation, intense isolated attosecond pulses with durations (photon energy) of 26 as (314 eV), 27 as (323 eV), 29 as (543 eV) and 27 as (574 eV) can be achieved for time delays $\tau$ = 0.41T, 0.51T, 1.21T and 1.31T, respectively. Moreover, the intensity of attosecond pulses has been increased up to one order of magnitude compared to the scenario without any phase compensation. 

\begin{figure}[t]
\centering\includegraphics[width=0.65\columnwidth]{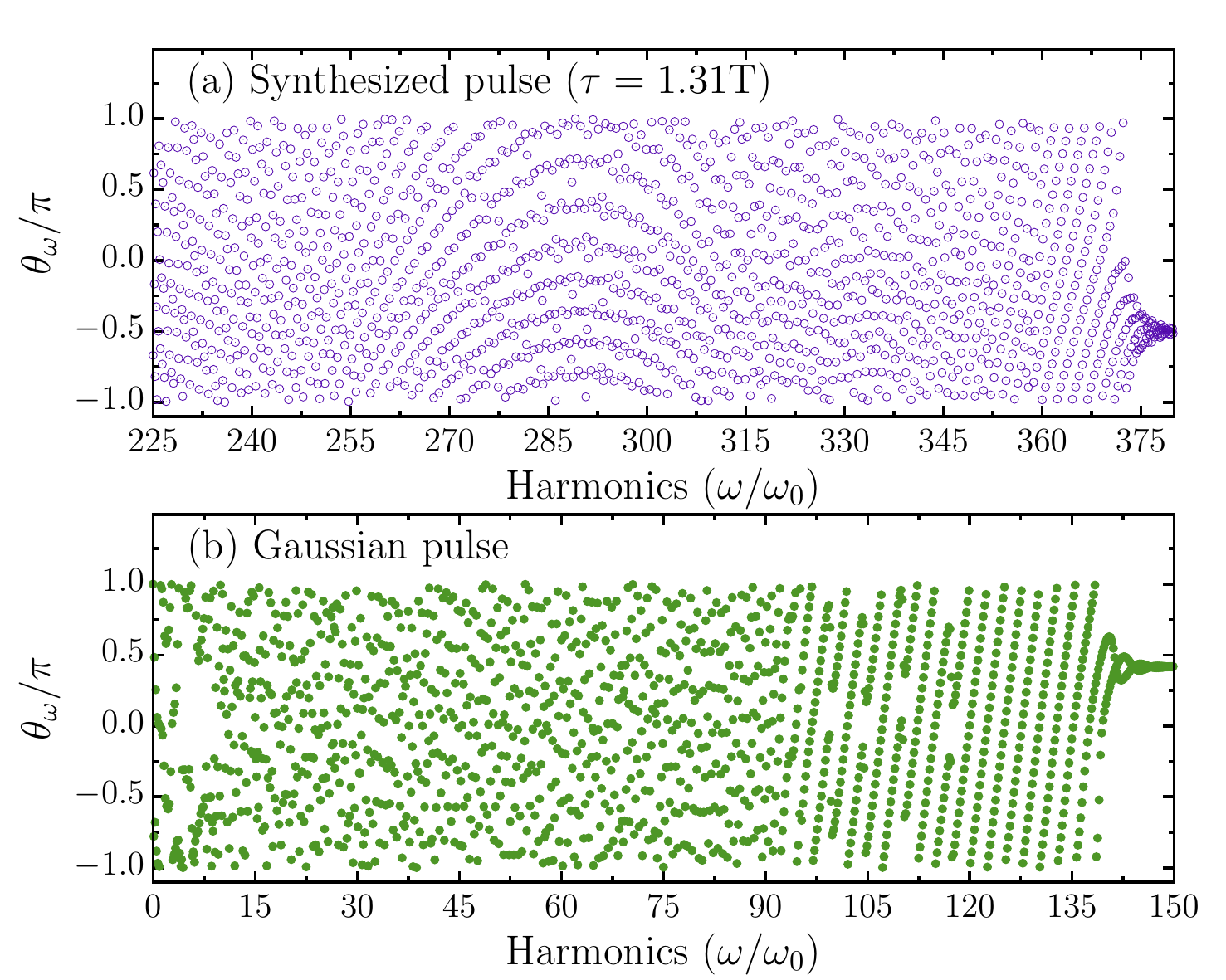}
\caption{Phase of emitted harmonics in case of : (a) Synthesized pulse with delay $\tau$ = 1.31T and (b) Gaussian-envelope pulse with laser parameters same as taken in figure  \ref{GaussCompare}(a).}
\label{Phase}
\end{figure}

One can notice that despite having a large harmonic continuum in the case of the synthesized field, the generated attosecond pulse is not as short as it is in the case of a Gaussian pulse \cite{Lu2009_JPhysB}. This can be understood by taking into account the variation of the phases of the emitted harmonics.  The phase of the emitted harmonic is calculated as: 
\be \theta_\omega = \text{arctan2}\Big(\text{Im}[D_\omega], \text{Re}[D_\omega] \Big),\ee
where, $D_\omega$ is the fourier transform of the dipole acceleration as calculated in equation(\ref{dipacc}). 
A comparison between the emitted harmonic phases for synthesized and Gaussian pulse is depicted in figure  \ref{Phase}(a) and \ref{Phase}(b), respectively. We see that, for the synthesized pulse, the previous $15$ harmonics from the cutoff (between 360th - 375th order) have constant phase variation between consecutive harmonics; while for the Gaussian pulse, nearly $50$ harmonics (between 90th - 140th order) are separated by a constant phase. The superposition of these phase-locked harmonics can generate an ASP of shorter duration without applying any phase compensation technique. However, if the phase dispersion is properly compensated, then the obtained large HHG continuum in case of the synthesized pulse can be used to produce a shorter ASP.
 
 \section{Strong Field Approximation} \label{SFA}  

So far we have discussed how the harmonic cutoffs can be regulated by simply changing the delay between the 
two oppositely polarized laser pulses. The cutoff scaling is found to be a property of the laser pulse envelope, and how 
it changes with the delay parameter. In order to further validate the scaling laws, in this section we study the high harmonic generation process under the framework of strong field approximation (SFA). The main assumption made in SFA is that the dynamics of the electron after ionization is solely controlled by the strong laser field and the effect of the core potential is small enough to be ignored. Within the regime of tunnel ionization the Keldysh parameter  $\gamma\ll 1$  \cite{Keldysh1965_JETP}, and the time dependent wavefunction $\ket{\psi(x,t)}$ of electron can be written as (dropping the $x$-dependence):
\be \ket{\psi(t)} = a(t) \ket{g} + \ket{\phi(t)} \ee 
where, $\ket{g}$ is the ground state which is calculated by the imaginary-time propagation method \cite{Bader2013_JChemPhys}, $\ket{\phi(t)}$ represents the continuum part of the wavefunction and $a(t)$ is the amplitude of the ground state and can be computed using the quasistatic approximation $| a(t) |^2 =\exp[-\int_0^t w(t') dt']$, where $w(t)$ is the instantaneous ionization rate calculated by the Ammosov-Delone-Krainov formula \cite{ADK1986_JETP}.

The continuum part $\ket{\phi(t)}$ under SFA satisfies the equation \cite{Lewenstein1994_PRA}:
\be i \ket{\dot{\phi}} = H_{V}(t) \ket{\phi} - E(t) x a(t) \ket{g}, \label{volkov eqn1} \ee
where $H_{V} \equiv -\frac{1}{2} \frac{\partial^2}{\partial x} - E(t) x + I_p$ is the Volkov Hamiltonian, with $I_p$ being the ionization potential of the atomic specie under study. The exact solution of equation(\ref{volkov eqn1}) can be given as \cite{Lewenstein1994_PRA}: 
\be \braket{p_x + A(t)| \phi}  = -i \int_0^t dt' a(t') E(t') \braket{ p_x + A(t')|x| g}  \text{e}^{-i S(p_x, t, t')},\label{volkov eqn2} \ee
where the phase factor 
\be S(p_x,t,t') \equiv \int_{t'}^t \frac{1}{2} (p_x + A(t''))^2 dt'' + I_p(t-t')\ee 
is the quasi-classical action and $\ket{p_x}$ denotes the plane wave state
\be \braket{x|p_x} = \frac{1}{\sqrt{2\pi}} \text{e}^{i p_x x}. \ee
Also, $A(t) = -\int_0^t E(t') dt'$ is the vector potential of the laser field $E(t)$.

The expectation value of the time dependent dipole acceleration $\ddot{d}(t)$ defined in equation(\ref{dipacc}) is:
\be
\ddot{d}_{SFA}(t) = -\braket{ \psi(t) | V'(x) | \psi(t)} + E(t)
 = -\big [a^{*}(t) \bra{ g} + \bra{\phi(t)} \big ] | V'(x) | \big [ a(t) \ket{g} + \ket{\phi(t)} \big ]
 + E(t).
\ee
Considering the continuum-bound transitions while dropping the $E(t)$ and higher order continuum-continuum transitions, the dipole acceleration can be written as:
\be \ddot{d}_{SFA}(t) = \ddot{\zeta}(t) + \ddot{\zeta}^*(t), \label{dipacc2}\ee where, 
\be \ddot{\zeta}(t) = -a^{*}(t) \braket{ g | V'(x) |\phi(t)}. \label{dipacc3}\ee
  
Now in order to calculate $\ddot{\zeta} $, we insert the momentum space completeness relation ($\int \ket{p_x +A(t)}\bra{p_x +A(t)} dp_x $) in equation(\ref{dipacc3}) and using equation(\ref{volkov eqn2}), we get:
\be
\ddot{\zeta}(t) = i \int_0^t dt' \int dp_x \braket{g| V'(x) | p_x + A(t)} a^{*}(t) a(t')\\
\times E(t') \braket{p_x + A(t') |x| g} \text{e}^{-i S(p_x, t, t')},
\label{dipacc4}
\ee
Applying the saddle point approximation to the momentum integration as in \cite{Le2016_JPB}, we obtain the dipole acceleration as:
\begin{multline}
\ddot{\zeta}(t) = -\int_0^t dt' \Big(\frac{2\pi i}{t -t' -i \epsilon} \Big)^{3/2}  \braket{g| V'(x) | p_{xs} + A(t)} a^{*}(t)\\ \times a(t') E(t') \braket{p_{xs} + A(t') |x| g} \text{e}^{-i S(p_{xs}, t, t')}.
\label{dipacc5}
\end{multline}
Here the infinitesimal $\epsilon$ comes from the regularized Gaussian integral around the stationary phase point and the value of saddle point $p_{xs}$ is given by:
\be p_{xs} = - \frac{1}{t-t'} \int_{t'}^t A(t'') dt''. \ee
Considering the case of He-atom, the term in equation(\ref{dipacc4}) representing the ionization from the ground state to continuum can  adequately be represented as \cite{Chang2016}:
\be E(t') \braket{p_{xs} + A(t') |x| g} \approx  -\frac{(2I_p)^{1/4}}{|E(t')|} \sqrt{\frac{w(t')}{\pi}} \ee  
Consequently the expression for dipole acceleration is then given by,
\be
\ddot{\zeta}(t) =\pi (2I_p)^{1/4} \int_0^t dt' \Big(\frac{2i}{t -t' -i \epsilon} \Big)^{3/2}  \braket{g| V'(x) | p_{xs} + A(t)} a^{*}(t) a(t') \frac{\sqrt{w(t')}}{|E(t')|} \text{e}^{-i S(p_{xs}, t, t')}.
\label{dipacc5}
\ee

Finally, similar to equation(\ref{eq:hhg}) the harmonic spectrum is obtained by doing the Fourier transformation of the dipole acceleration $\ddot{d}_{SFA}(t)$, i.e., 
\begin{equation}
S(\omega) = \Big|\frac{1}{\sqrt{2\pi}} \int \ddot{d}_{SFA}(t)\exp{[-i\omega t]} dt \Big|^2.
 \label{eq:hhg2}
\end{equation}

 \begin{figure}[t]
\centering\includegraphics[width=0.65\columnwidth]{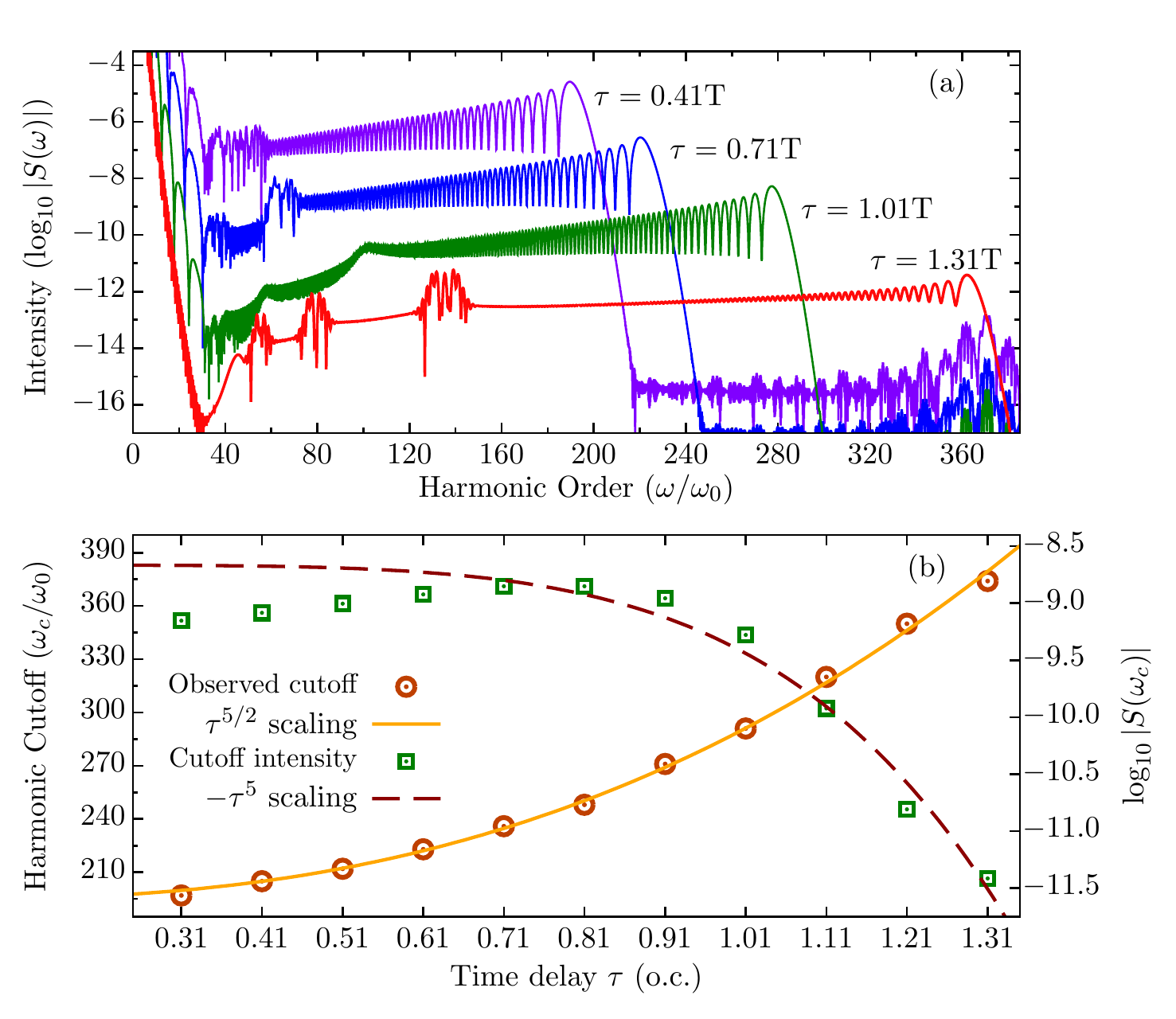}
\caption{ Under the SFA framework, higher harmonic spectra for He atom is presented for different delay parameters (a). The harmonic cutoffs and intensity scaling are also presented (b). All the laser parameters are same as used in figure  \ref{Scale}. }
\label{sfa:result}
\end{figure} 

\begin{table}[b]
\caption{\label{table2}The summary of the scaling parameters is presented. The harmonic cutoffs are fitted through $\omega_c(\tau) = \mu \tau^{5/2} + \eta$ and cutoff intensities are fitted through $I_c(\tau) = - \delta\tau^\kappa + \chi$. Here,  the intensity scaling parameters $\delta$ and $\chi$ are normalized with respect to the corresponding values as obtained for $\tau = 0.71$T case.} 
\footnotesize
\begin{indented}
\item[]\begin{tabular}{@{}llllll}
\br
Approach &  $\ \ \mu$  & $\ \ \eta$  &  $\ \delta$ & $\kappa$ & $\ \ \ \chi$ \\  
\mr
Classical trajectories     & 94.4    & 194.3    &  -  &  - &  -    		   \\  
He, Soft-core potential    & 94.4    & 194.3    & 0.18    & 5.0   & -0.97  \\  
He, Pliant-core potential  & 94.4    & 194.3    & 0.14    & 6.0   & -0.99  \\  
He, Soft-core potential, SFA & 94.0    & 194.8    & 0.08    & 5.0   & -0.98  \\  
H, Lower intensity         & 19.3    & 42.09    & 0.17    & 4.0   & -0.95  \\  
\br
\end{tabular}
\end{indented}
\end{table}
\normalsize
 
 Figure \ref{sfa:result} depicts the main results of our SFA-based simulations for He-atom. To compare the results with the scaling laws suggested by the 1D TDSE simulations, the SFA-based harmonic spectrum was calculated for the same laser pulse as in the case of 1D TDSE, i.e., for a synthesized laser pulse with varying $\tau$ values and fixed pulse energy. The harmonic spectrum for different time delays  is presented in figure  \ref{sfa:result}(a). The HHG spectra corresponding to $\tau$ = 0.41T, 0.71T and 0.91T are shifted vertically for clarity purposes. In comparison with the HHG spectra obtained from 1D TDSE solution and shown in figure  \ref{lg_vg}, one can see that the harmonic cutoffs are identical for corresponding delay cases. Moreover, the selective enhancement in intensity of harmonics at similar positions can also be seen. In figure  \ref{sfa:result}(b), we have presented the harmonic cutoffs ($\omega_c/\omega_0$) as obtained from the SFA-based simulations along with the fitted curve (solid-line), showing the same scaling of $\sim \tau^{5/2}$ as in the case of 1D TDSE [see figure \ref{Scale}(a)]. The intensities of the respective harmonic cutoff for different time delays are also shown in figure  \ref{sfa:result}(b), along with a reference curve (dashed-line) which is following a scaling of $\sim -\tau^5$. It can be seen that for $\tau > 0.71$T cases, the cutoff intensities ($\log_{10} |S(\omega_c)|$) are following the   $\sim -\tau^5$ scaling, supporting the results obtained from 1D TDSE simulations [refer figure  \ref{Scale}(a)]. However, for  $\tau < 0.71$ cases, the observed intensities of cutoff harmonic start differing from the scaling law with decreasing values of delay $\tau$. This can be understood by recalling the fact that we have taken the quasi-classical approximation in the SFA-based calculations and as mentioned earlier, decreasing the value of delay parameter will shorten the laser cycle resulting in lowering the effective wavelength of the synthesized pulse.  

\vspace{0.5cm}
\section{Summary and Conclusions}
\label{con}

We have theoretically investigated the higher harmonic generation and the generation of a single attosecond pulse from a He atom. The driving field is sculpted from a sinc-shaped pulse  by a simple experimentally realizable setup,  though, technological limitations currently restrict the generation of the sinc-shaped pulse at mentioned intensities. This synthesized field can be considered as a single cycle pulse which can favorably control the electron quantum path. It has been observed that, in comparison with a few-cycle Gaussian pulse of the same energy, the bandwidth of an XUV supercontinuum spectrum is significantly broadened for the synthesized pulse.  We find that the energy of the cutoff harmonics increases with the increasing delay parameter ($\tau$). Specifically, it nicely scales as $\sim\tau^{5/2}$ for a fixed driver pulse energy. Furthermore, the intensity of the cutoff harmonic is also seen  to be following a $\sim-\tau^{4\textnormal{-}6}$ scaling, depending on the atomic species or model potential under study. These well-defined scaling laws for the harmonic cutoff and its intensity indicate a realizable experimental setup, wherein radiations from XUV to soft-Xrays can be generated by simply changing the distance $d$ on an optical bench [refer figure  \ref{geo}].  The mentioned scaling laws for the harmonic cutoffs and the cutoff intensities are validated by (a) classical trajectory analysis [refer figure  \ref{clatraj}], (b) gauge invariant HHG spectra [refer figure  \ref{lg_vg}], (c) soft-core and pliant-core potentials [refer figure  \ref{Scale}], and (d) strong-field-approximations [refer figure  \ref{sfa:result}]. In Table \ref{table2} we summarize the scaling parameters as obtained through different approaches. 

We have used these higher harmonics to generate attosecond pulses having a central frequency in XUV to the soft-Xray regime of the electromagnetic spectrum. This has been achieved by filtering the harmonics of bandwidth $155$ eV from the respective harmonic cutoff for a given delay parameter. The quantum time-frequency analysis shows that the contribution of the harmonics corresponding to the shorter quantum path in the HHG supercontinuum spectrum decreases with the increasing delay in the pulse synthesis. As a result, an isolated $\sim 100$ as pulse for $\tau = 1.31$T is straightforwardly obtained without any phase compensation. If the phase is compensated correctly, then an intense ultrashort $\sim 27$ as pulse can be generated with photon energy $\sim 570$ eV. 

The advantage of the scheme proposed here lies in the tuning of broadband XUV supercontinuum by merely adjusting the time delay on an optical bench, which seems feasible from an experimental point of view. The harmonic cutoff scaling is found to be a property of the laser pulse, however the intensity scaling would depend on the atomic species or model potential under study.  The enhancement in some satellite harmonics is also studied, and its dependence on delay parameter $\tau$ has been summarized. Exploration of higher dimensional effects on the electron quantum path and the corresponding attosecond pulse generation using this synthesized sinc pulse is reserved for the future project.  

\section*{Acknowledgments}
Authors acknowledge the Science and Engineering Research Board, Department of Science and Technology, Government of India, for funding the project EMR/2016/002675. 

\section*{ORCID iDs}
Rambabu Rajpoot \orcidicon{0000-0002-2196-6133} \url{https://orcid.org/0000-0002-2196-6133}\\
Amol R. Holkundkar \orcidicon{0000-0003-3889-0910} \url{https://orcid.org/0000-0003-3889-0910}\\
Jayendra N. Bandyopadhyay \orcidicon{0000-0002-0825-9370} \url{https://orcid.org/0000-0002-0825-9370}
 
\footnotesize

\end{document}